\documentclass[11pt,titlepage]{article}
\usepackage{setspace}
\usepackage{epsfig}
\usepackage{ae}
\usepackage{aecompl}
\usepackage{amsmath}
\usepackage{subfig,graphicx}
\usepackage[round,numbers,sort&compress]{natbib}
\begin{document}
\title{The Effect of Solutes on the Temperature of Miscibility
  Transitions in Multi-component Membranes}
\author{ D. W. Allender$^{1,2}$ and   M. Schick$^1$\\
$^1$Department of Physics, 
University of Washington, Seattle, Washington\\
$^2$Department of Physics, Kent State University, Kent, Ohio}
\date{\today}
\maketitle
\begin{abstract} 
We address questions posed by experiments which show that most small-chain 
alcohols reduce the miscibility transition temperature when added to giant 
plasma membrane vesicles, but increase that temperature when added to 
giant unilamellar vesicles. In both systems the change in temperature 
depends non-monotonically on the length of the alcohol chain. To emphasize 
the roles played by the internal entropies of the components, we model 
them as linear polymers. We show that, within Flory-Huggins theory, the 
addition of alcohol causes an increase or decrease of the transition 
temperature depending upon the competition of two effects.  One is the 
dilution of the solvent interactions caused by the introduction of solute, 
which tends to lower the temperature. The other is the preference of the 
solute for one phase or the other, which tends to raise the temperature. 
The magnitude of this term depends on the entropies of all components. 
Lastly we provide a reasonable explanation for the behavior of the 
transition temperature with alcohol chain length observed in experiment.
  \end{abstract}

\section{Introduction}

\begin{figure}[htbp]    
\includegraphics[width=.5\columnwidth] {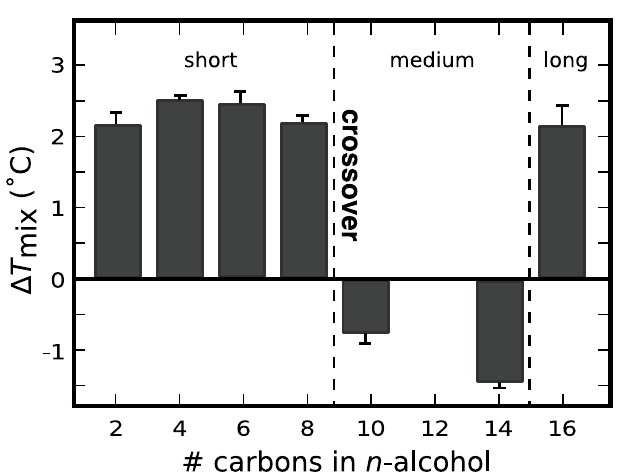}
\caption{ Change in transition temperature upon the addition of n-alcohol 
to a GUV comprised of mol fractions 35/35/30 DOPC/DPPC/cholesterol. 
Concentrations of solute are all three times a fiducial given in Ref. 
\cite{pringle81}. Figure from Ref. \cite{cornell17}.}
\label{caitlin}
\end{figure}

A long-accepted means of interrogating the properties of a bilayer 
membrane is to add to it a molecule whose properties are well understood. 
In this spirit, a series of experiments were carried out in which 
short-chain alcohols were introduced into giant plasma membrane vesicles 
(GPMVs)\cite{gray13,machta16}. It was found that the addition of most of 
the alcohols caused a decrease in the temperature at which the system 
transitioned from a single uniform phase to two coexisting liquid-ordered 
(lo) and liquid-disordered (ld) phases. However, when these same alcohols 
were introduced into giant unilamellar vesicles (GUVs) composed of a 
ternary mixture of di\-oleoyl\-phosphatidylcholine (DOPC), 
di\-palmitoyl\-phosphatidylcholine (DPPC) and cholesterol, the transition 
temperature usually increased\cite{cornell16,cornell17}.  These two quite 
disparate results did share one feature; the behavior of the transition 
temperature changed non-monotonically as a function of the length of the 
$n$-alcohol chain. In the GPMVs, addition of n-alcohols with $2\leq n\leq 
10$ reduced the transition temperature \cite{gray13} while hexadecanol 
increased it \cite{machta16}. In the GUVs, as shown in Fig. \ref{caitlin}, 
alcohols with $2\leq n\leq 8$ increased the transition temperature, those 
with $10\leq n \leq 14$ decreased it, and hexadecanol increased it once 
again. We want to understand the difference in behavior between the two 
systems and the origin of the non-monotonic behavior of the transition 
temperature with the chain length of the alcohol.

The problem of predicting the effects of a solute on a solvent which can 
undergo a phase transition is a rather old one. Over fifty years ago, 
Prigogine and Defay devoted a chapter to it in their volume ``Chemical 
Thermodynamics'' \cite{prigogine54}. By examining a regular solution of 
structureless components they concluded (a) ``Hence the introduction of a 
third component which is equally soluble in the first two components will 
lower the critical solution temperature...'' and (b) ``Hence the addition 
of a third component which is much less soluble in one of the first two 
components than in the other, will always raise the critical solution 
temperature...''.

  Recently these statements were made more quantitative by the
  simulation of a simple Ising model \cite{meerschaert15}.  In it, the
  ``solvent''  spins interacted with their nearest neighbors with a
  strength $J$, while a small amount of a third, ``solute'',  
spin interacted with its neighbors with strength $gJ$.  
It was found that the transition temperature increased if $|g|>1$, and 
decreased if $|g|<1$. The lowest transition temperature occurred when the 
third component interacts with the other two equally, i.e. when $g=0.$ The 
conditions which bring about the lowest transition temperature in a 
membrane are of interest because the transition temperature scales with 
the effective interaction between lipids. Furthermore many other 
properties of the membrane scale with this interaction, such as its 
internal pressure profile. The circumstances which bring about the lowest 
transition temperature, then, correspond to a membrane in which this 
interaction scale is reduced by the greatest amount.

Because of the simplicity of the Ising model, the behavior of the 
transition temperature depends upon a single parameter, $g$. Because of 
the symmetry of the Ising model, which corresponds to the equality of the 
chemical potentials of the two major components at the same temperature, 
the entropy of the two coexisting phases are equal. The lack of internal 
entropy of the components, and the equality of the entropy of the 
coexisting phases are not characteristic of biological lipid components 
nor of their phases \cite{portet12,schick16}. That the difference in 
entropy of the coexisting phases is of importance was pointed out by one 
of us \cite{schick16} who noted that it is well-known \cite{landau58} that 
the temperature of a transition in a one-component system is decreased if 
a solute partitions preferentially into the phase with greater entropy. It
was then shown that the tendency remains in a multi-component membrane. 
However the difference in the behavior of the 
transition temperature in GUVs and GPMVs was not explained.

It is the purpose of this paper to illuminate the effect of the internal entropies of membrane components and of solute on the behavior of a miscibility transition temperature as solute is added. We do this by considering 
 membrane components $A$ and $B$ as polymer chains of polymerization 
indices $N_A$ and $N_B$, and the solute as a polymer chain of index $N_S$. 
The critical temperature is calculated within mean-field theory as a 
function of the solute volume fraction. For small volume fractions, 
comparable to those in the experiments noted above, we find that the 
behavior of the miscibility transition temperature as a function of solute 
volume fraction results from a competition between two tendencies: the 
first is simply that the introduction of any solute reduces the number of 
solvent interactions and therefore lowers the transition temperature. The 
second tendency is essentially that noted by Prigogine and Defay:  if the 
solute prefers one phase to the other, that will tend to increase the 
transition temperature. However the magnitude of this increase depends not 
only upon the difference in the interactions between the solute and the 
solvent components, but also upon the intrinsic entropies of the membrane 
components and of the solute. This term, in general, is not minimum when 
the solute partitions equally into the two phases, but rather when it 
partitions somewhat preferentially into the phase with the larger entropy. 
With the aid of our results, we are able to understand the difference in 
the behavior of the transition temperature in GPMVs and GUVs. Further, by 
combining them with previous results for the partitioning of n-alcohols in 
lo and ld phases \cite{uline12}, we are able to provide an explanation for 
the non-monotonic behavior with chain length of the miscibility transition 
temperature in membranes containing $n$-alcohols.
\section{Methods}
\subsection{The Model}

We consider an incompressible membrane at temperature $T$ comprised of 
$n_A$ molecules of component $A$ and $n_B$ molecules of component $B$. In 
order to consider the effects of internal entropy, we treat the components 
as linear polymers with polymerization indices $N_A$ and $N_B.$ To this 
membrane, we add $n_S$ molecules of a solute, also treated as a linear 
polymer with polymerization index $N_S$. Because the system is 
incompressible, its volume, $\Omega$, is not a thermodynamically 
independent variable, but is related to the number of molecules of the 
components according to
\begin{equation}
\label{volume}
\Omega(n_A,n_B,n_S)=v_0(n_AN_A+n_BN_B+n_SN_S),
\end{equation}
where $v_0$ is the monomer volume of $A$, $B$, or $S$ chains, volumes 
which are assumed to be equal.

In mean-field, or Flory-Huggins, approximation \cite{degennes79}, the 
Helmholtz free energy of the system, $F$, can be written
\begin{eqnarray}
\label{helmholtz}
&&F(T,n_A,n_B,n_S)=\nonumber\\
&&\qquad V_{AB}\ n_AN_A\frac{n_BN_Bv_0}{\Omega}+ V_{AS}\ n_AN_A\frac{n_SN_Sv_0}{\Omega} 
+V_{BS}\ n_BN_B\frac{n_SN_Sv_0}{\Omega}\nonumber\\
&&\ +k_BT\left\{n_A\ln\left[\frac{n_AN_Av_0}{\Omega}\right]+ 
n_B\ln\left[\frac{n_BN_Bv_0}{\Omega}\right]+n_S\ln\left[\frac{n_SN_Sv_0}{\Omega}\right]\right\}
\end{eqnarray}
where $k_B$ is Boltzmann's constant.  The energy $V_{AB}$ is related to the interaction energy between pairs of $A$ monomers, ${\tilde V}_{AA}$, pairs of $B$ monomers, ${\tilde V}_{BB}$, and $AB$ pairs, ${\tilde V}_{AB}$,  according to
\begin{equation}V_{AB}={\tilde V}_{AB}-\frac{1}{2}({\tilde V}_{AA}+{\tilde V}_{BB}),
\end{equation}
and similarly for $V_{AS}$ and $V_{BS}$. From the free energy, the 
chemical potentials of the three components are obtained
\begin{equation}
	\mu_{\alpha}=\frac{\partial F(T,n_A,n_B,n_S)}{\partial n_{\alpha}},\qquad \alpha=A,B,S.
\end{equation}
The chemical potentials are functions of temperature and two other intensive quantities. It is convenient to take them to be the volume fractions of the $A$ and $S$ components
\begin{eqnarray}
\label{fractions}
\Phi_A&\equiv&\frac{n_AN_Av_0}{\Omega},\\
\Phi_S&\equiv&\frac{n_SN_Sv_0}{\Omega}.
\end{eqnarray}
 The volume fraction of the $B$ component is then $\Phi_B=1-\Phi_A-\Phi_S.$

\subsection{Method of Solution}

Given a net repulsive interaction, $V_{AB}>0$, there will be a transition 
from one uniform phase to two coexisting phases, $I$ and $II$, below some 
critical temperature. The equations which determine the volume fractions 
of the components at coexistence are
\begin{eqnarray}
\label{chempot}
\mu_A(T,\Phi_A^I,\Phi_S^I)=\mu_A(T,\Phi_A^{II},\Phi_S^{II}),
\nonumber\\
 \mu_B(T,\Phi_A^I,\Phi_S^I)=\mu_B(T,\Phi_A^{II},\Phi_S^{II}),
\nonumber\\
\mu_S(T,\Phi_A^I,\Phi_S^I)=\mu_S(T,\Phi_A^{II},\Phi_S^{II}).
\end{eqnarray}
These three equations in five unknowns determine the surface of 
coexistence $T(\Phi^I_A,\Phi^I_S), \Phi^{II}_A(\Phi^I_A,\Phi^I_S)$ and 
$\Phi^{II}_S(\Phi^I_A,\Phi^I_S).$

The line of critical compositions, $\Phi_{A,c}=\Phi_{A,c}(\Phi_S)$, is 
obtained from the non-trivial solution of the two homogeneous equations in 
two unknowns
\begin{eqnarray}
\label{homo}
\Phi^{II}_A(\Phi^I_A,\Phi^I_S)-\Phi^I_A&=&0,\\
\label{homo2}
\Phi^{II}_S(\Phi^I_A,\Phi^I_S)-\Phi^I_S&=&0.
\end{eqnarray}
From this line of critical compositions, the critical temperature is 
obtained as a function of solute volume fraction, 
$T_c(\Phi_S)=T(\Phi_{A,c}(\Phi_S),\Phi_S).$

To obtain the properties of the critical line, we expand Eqs. (\ref{chempot})
in the small parameters $\Phi^{II}_A-\Phi^I_A$ and
$\Phi^{II}_S-\Phi^I_S$. If we keep only terms in linear order, the
resulting equations are not independent, but yield 
the two linear homogeneous equations, Eqs. (\ref{homo}) and (\ref{homo2}), noted above.
Setting the determinant of these equations to zero, we obtain 
the following expression for the critical temperature in terms of the unknown
critical volume fraction, $\Phi_{A,c}$, and the given solute volume
fraction $\Phi_s$;

\begin{eqnarray}
k_BT_c&=&V_{AB}\frac{\beta+(\beta^2+4\alpha\gamma)^{1/2}}{2\alpha},\qquad
{\rm where}\nonumber\\
\alpha&\equiv&\frac{1}{N_A\Phi_{A,c}}\left[1+\frac{N_S\Phi_S}{N_B(1-\Phi_S)}\right]+\frac{1}{N_B(1-\Phi_{A,c}-\Phi_S)}\left[1+\frac{N_S\Phi_S}{N_A(1-\Phi_S)}\right],\nonumber\\
\beta&\equiv&2\left[1+\frac{N_S\Phi_SV_{BS}}{N_A\Phi_{A,c}V_{AB}}+\frac{N_S\Phi_SV_{AS}}{N_B(1-\Phi_{A,c}-\Phi_S)V_{AB}}\right]\nonumber,\\
\gamma&\equiv&N_S\Phi_S\left[1-2\frac{(V_{AS}+V_{BS})}{V_{AB}}+\left(\frac{V_{AS}-V_{BS}}{V_{AB}}\right)^2\right].
\end{eqnarray}
We also obtain from the non-trivial solution of the homogeneous
equations the ratio 
$(\Phi_S^{II}-\Phi_S^I)/(\Phi_A^{II}-\Phi_A^I)\equiv M,$ which is the
slope of the tie-line. It is evaluated at the critical point, $M_c$, and
is also
expressed in terms of the unknown critical volume fraction, $\Phi_{A,c}$ and
the given solute volume fraction $\Phi_S$; 
\begin{equation}
M_c=-N_S\Phi_S\frac{k_BT_c-N_B(1-\Phi_{A,c}-\Phi_S)(V_{AB}+V_{BS}-V_{AS})}{[k_BT_c-2N_S\Phi_SV_{BS}]N_B(1-\Phi_{A,c}-\Phi_S)+k_BT_cN_S\Phi_S}.
\end{equation}
Finally, if we include in our expansion of Eqs. (\ref{chempot}) terms of
third order in the small expansion parameters, we recover a third
independent equation  
\begin{eqnarray}
0&=&\frac{1}{N_A\Phi_{A,c}^2}-\frac{(1+M_c)^3}{N_B(1-\Phi_{A,c}-\Phi_S)^2}+\frac{M_c^3}{N_S\Phi_S^2}.
\end{eqnarray}
This, with the other two equations above, determines
$\Phi_{A,c}(\Phi_s)$ and $M_c(\Phi_S)$, and $T_c(\Phi_s)$.

\section{Results}
We first consider the critical concentration, $\Phi_{A,c}(0),$ and critical 
temperature, $T_c(0)$, in the limit of no solute, $\Phi_S\rightarrow 0.$ 
Following the above procedure, we obtain for the critical concentration of 
the $A$ component the result \cite{degennes79}
\begin{equation}
\Phi_{A,c}(0)=\frac{\sqrt{N_B}}{\sqrt{N_A}+\sqrt{N_B}},
\end{equation}
 and find the critical temperature to be given by
\begin{equation}
2\frac{V_{AB}}{k_BT_c(0)}=\left[\frac{1}{\sqrt{N_A}}+\frac{1}{\sqrt{N_B}}\right]^2.
\end{equation}
In the symmetric case for which $N_A=N_B\equiv N$, these results reduce to $\Phi_A=\Phi_B=1/2$ and $NV_{AB}/k_BT=2$. (In the polymer literature, $V_{AB}/k_BT$ is denoted $\chi$ so that this relation  is written $\chi_cN=2$, a well-known result\cite{degennes79}.)

Now let the solute be introduced. With the interactions between solute and 
the solvent components non-zero, we find it convenient to characterize 
them by the average interaction, normalized by $V_{AB}$ and the difference 
in the interactions, again normalized;
\begin{eqnarray}
r\equiv \frac{V_{AS}+V_{BS}}{2V_{AB}},\nonumber\\
\delta r\equiv \frac{V_{AS}-V_{BS}}{V_{AB}}.
\end{eqnarray}
We now solve the equations for the critical temperature in a power series 
in the solute volume fraction. Introducing
\begin{eqnarray}
\epsilon&\equiv&\sqrt{N_A}+\sqrt{N_B},\nonumber\\
\nu&\equiv&\sqrt{N_A}-\sqrt{N_B},
\end{eqnarray}
we obtain
\begin{eqnarray}
\label{mc}
   M_c&=&-\frac{N_S}{2N_AN_B}\epsilon (\nu+\epsilon\delta r)\Phi_S+O(\Phi_S^2),\\
   \label{phic}
   \frac{\Phi_{A,c}(\Phi_S)-\Phi_{A,c}(0)}{\Phi_{A,c}(0)}&=&
   -\left[1-N_s(\nu+\epsilon\delta r)\left(3-(\nu+\epsilon\delta r)^2\frac{N_S}{4N_AN_B}\right)\frac{N_A^{1/2}}{4N_AN_B}\right]\Phi_s\nonumber\\
   &&+O(\Phi_S^2),
\end{eqnarray}
\begin{eqnarray}
\label{tshift}
\frac{T_c(\Phi_S)-T_c(0)}{T_c(0)}&=&c_1\Phi_S+c_2\Phi_S^2+O(\Phi_S^3),\nonumber\\
c_1&=&-1+\frac{N_S}{4N_AN_B}(\nu+\epsilon\delta r)^2,\nonumber\\
c_2&=&-\left[\frac{N_S}{4N_AN_B}(\nu+\epsilon\delta r)\right]^2\{[\epsilon^2(1-4r+(\delta r)^2]\nonumber\\&+&(N_AN_B)^{1/2}(c_1)^2\}.
\label{result1}
\end{eqnarray}
Before discussing the effect of the solute on the transition temperature, 
we simply note the following: if $\delta r>0$, the solute prefers the 
B-rich phase, hence a negative slope of the tie-line, as defined above, is 
expected at the critical point. That this is the case is shown by Eq. 
(\ref{mc}). The change in the volume fraction of component $A$ at 
criticality is a result of two effects. The first term in Eq. (\ref{phic}) 
is simply the effect of dilution; that is, as a volume fraction of solute 
is introduced, the volume fraction of the other components must decrease.  
The second term shows that if the solute prefers the B-rich phase, this 
will tend to increase the volume fraction of the $A$ component at the 
critical point. Note that the presence of the factor $\nu$ expresses a 
purely entropic effect.

For the shift in the critical temperature due to the addition of solute, 
which is the principal result of this paper, we note that for the 
millimolar concentrations employed in the experiments of interest, it is 
sufficient to keep in Eq. (\ref{tshift}) only the term linear in $\Phi_S$,
\begin{equation}
\label{result}
\frac{T_c(\Phi_S)-T_c(0)}{T_c(0)}=\left[-1+\frac{N_S}{4N_AN_B}(\nu+\epsilon\delta r)^2\right]\Phi_S.
\end{equation}
From thisexpression, one sees that the effect on the temperature at criticality, 
just as on the concentration of component $A$ at criticality, results from a 
competition between two terms. The first, $-\Phi_S$, is simply the 
reduction in the transition temperature due to the dilution of the number 
of $AB$ interactions resulting from the introduction of the solute. This 
effect would be present no matter the nature of the solute. The second 
term, being positive, always tends to increase the transition temperature. 
It reflects, but modifies, the dictum of Prigogine and Defay 
\cite{prigogine54} that ``...the addition of a third component which is 
much less soluble in one of the first two components than in the other, 
will always raise the critical solution temperature...''. Indeed if 
$\delta r=(V_{AS}-V_{BS})/V_{AB}$ is large in magnitude, and of either 
sign, then this term will tend to increase the transition temperature. 
Whether it will actually do so or not depends upon the intrinsic entropies 
of the solute as well as those of the solvent, $A$ and $B$. It is 
interesting that this effect depends only on the magnitude of the 
difference, $\delta r$, between the solute interactions; it is independent 
of the strength of these interactions, encapsulated in the parameter $r$. 
Furthermore we see that the largest reduction of the transition 
temperature does not occur when the solute interacts equally with the 
solvent components $A$ and $B$, i.e. $\delta r=0$, but rather when there 
is a difference in its interactions equal to
\begin{eqnarray}
\delta r&\equiv&\frac{V_{AS}-V_{BS}}{V_{AB}}\nonumber\\
&=&-\frac{\nu}{\epsilon}
  =-\frac{\sqrt{N_A}-\sqrt{N_B}}{\sqrt{N_A}+\sqrt{N_B}}.
\end{eqnarray}
Recalling that a positive interaction is repulsive, we see that if 
$N_A>N_B$, then the above states that the greatest reduction in transition 
temperature occurs when the solute partitions preferentially into the 
$A$-rich phase. As we show below, near the critical transition, this phase 
is the one with larger entropy per unit volume. Hence this nicely 
illustrates the point made in Ref. \cite{schick16} that if the solute 
partitions preferentially into the phase with greater entropy, the 
transition temperature of the multi-component system will tend to 
decrease. Again, whether it actually does so depends on the magnitude of 
this term compared to unity.

To see that the entropy per unit volume of the $A$-rich phase, phase $I$, 
is greater than that of the $B$-rich phase, phase $II$, near the critical 
point, if $N_A>N_B,$ we note that, from Eqs. (\ref{helmholtz}) and 
(\ref{fractions}), in the absence of solute, the entropy per unit volume, 
$S/\Omega$, is given by
\begin{equation}
\frac{S v_0}{\Omega k_B}=-\left[\frac{\phi_A}{N_A}\ln(\phi_A)+\frac{(1-\phi_A)}{N_B}\ln(1-\phi_A)\right]
\end{equation}
The volume fraction $\Phi_A$ in the $A$-rich phase can be written 
$\Phi_A=\Phi_{A,c}+\delta \Phi_A$ with $\delta \Phi_A>0.$ Near the 
critical point, $\delta \Phi_A/\Phi_{A,c}$ is small, so the above can be 
expanded in this parameter with the result
\begin{equation}
\frac{(S^I-S^{II})v_0}{\Omega k_B}\approx-2\left\{\frac{\ln(\phi_{A,c})+1}{N_A}-\frac{\ln(1-\phi_{A,c})+1}{N_B}\right\}\delta \Phi_A
\end{equation} 
Finally setting $N_A=N+\delta N$, $N_B=N-\delta N$ with $\delta N>0$ and expanding in $\delta N/N$, we obtain
\begin{equation}
\frac{(S^I-S^{II})v_0}{\Omega k_B}\approx2(3-2\ln 2)\frac{\delta N}{N^2}\delta\Phi_A\approx 3\frac{\delta N}{N^2}\delta\Phi_A
\end{equation}
which is positive.

\section{Discussion}
As we have seen, the resolution of the question of whether the addition of 
a solute to a solvent that can undergo a miscibility phase transition will 
raise or lower the transition temperature depends on the competition 
between two effects. The first, which is independent of the nature of the 
solute, is simply that the introduction of a solute reduces the number of 
interactions between solvent components and therefore tends to lower the 
transition temperature. The second depends on the nature of the solute. If 
it strongly favors one phase or the other, its introduction tends to raise 
the transition temperature. The magnitude of this second effect, and 
therefore the result of the competition between terms, depends upon the 
internal entropies of solvent and solute components. The behavior of the 
transition temperature is neatly encapsulated in the behavior of the 
quantity $c_1$ in Eqs. (\ref{result1}) and (\ref{result}),
\begin{equation}
\label{tada}
   c_1\equiv -1 +\frac{N_S}{4N_AN_B}(\nu+\epsilon\delta r)^2.
   \end{equation}
  If it is positive, the transition temperature increases, if negative the 
transition temperature decreases. We can understand from this how it can 
come about that the addition of short-chain alcohols to GPMVs causes a 
decrease in the transition temperature while the same alcohols in GUVs can 
cause an increase. First, the GPMVs contain long chain poly-unsaturated 
fatty acids which the GUVs do not. As a consequence the effective chain 
length, $N_A,$ of the component which is more abundant in the 
liquid-disordered phase is larger in GPMVs than it is in GUVs.  This 
decreases the prefactor of the second term in the expression for the
shift in transition temperature,  Eq. (\ref{result}), thereby
diminishing the effect of the interactions and making a decrease 
in the transition temperature more likely.  
This is an entropic effect. Second, it can be 
argued \cite{cornell17} that the difference in the order between the lo 
and ld phases is less in GPMVs than in GUVs, hence the difference in 
interactions, $\delta r,$ between solute and solvent components is less. 
This also makes a reduction in the transition temperature more likely.
 
\begin{figure}[htbp]    
\includegraphics[width=.5\columnwidth] {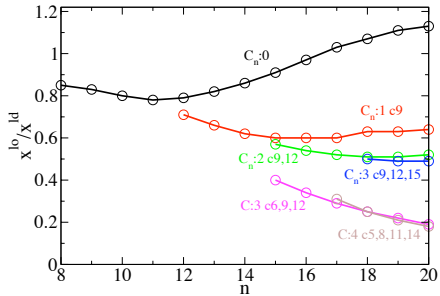}
\caption{Partition coefficient, $X^{lo}/X^{ld}$ for several kinds of single chains of length $n$. $C_n:0$ denotes a chain of length $n$ and no double bonds, while $C_n:1\ c9$ denotes a chain of length $n$ with one double bond at the ninth position, etc. From Ref. \cite{uline10}.}
\label{partition}
\end{figure}

To understand the variation in $T_c(\Phi_S)$ with the length of the 
n-alcohol chain, we must understand the behavior of the partitioning of 
the alcohol into the two phases, lo and ld, as a function of $n$. 
Fortunately the partition coefficient, the ratio of the fraction of 
alcohol in the lo phase to that in the ld phase, has been calculated for a 
series of acyl chains in a membrane composed of DPPC, DOPC, and 
cholesterol \cite{uline10}. The results for saturated chains and various 
unsaturated chains are shown in Fig. \ref{partition}.
    
 From the figure, we see that short-chain n-alcohols with $n<14$ partition 
preferentially into the ld phase, from which we infer $\delta r$  is negative. Presumably this 
preference is due to the area per 
lipid head group being  larger in that phase. For these values of $n$, we do 
not expect the magnitude of $\delta r$ to be small. With increasing $n,$ 
the energy penalty of repulsive interactions between the alcohol chain and 
the DOPC chain due to its double bond increases.  In addition the 
favorable energy of interaction with the ordered chains of the DPPC 
increases. Insertion of the alcohol into the lo phase is opposed, however, 
by the cholesterol. As a consequence of these various factors, there will 
be a range of $n$ for which the alcohol partitions roughly equally into 
the two phases. In this range $\delta r$ is small.  Eventually for 
sufficiently large $n$, on the order of 16 to 18, the n-alcohol partitions 
predominantly into the lo phase.  For such values of $n$, $\delta r$ is 
once again not small, but now is positive. Based on these observations and our analysis 
encapsulated in Eq. (\ref{tada}) we can predict that, (i), even in the GUVs, 
for which most alcohols increase the transition temperature, there will be 
an interval of $n$ over which the transition temperature decreases on the 
addition of $n$-alcohols. That this is indeed the case is seen in Fig. 
\ref{caitlin}.  Furthermore because of the presence of the entropic term 
$\nu=\sqrt{N_A}-\sqrt{N_B}>0$ in Eq. (\ref{tada}), we would predict, (ii), 
that the onset of the interval in $n$ over which the transition 
temperature decreases in the GUVs will occur when the alcohol still 
partitions more favorably into the ld phase. That this is correct is seen 
by a comparison of Figs. \ref{caitlin} and \ref{partition}.

There are other predictions that can be made on the basis of Eq. (\ref{tada}). 
For example, we expect that the change in transition temperature would be 
larger in membranes composed of shorter lipids than in those of longer 
ones, and that the effect will be larger were chains with double bonds to 
be added instead of the saturated alcohols. Of course any molecule that 
strongly prefers one phase to the other will tend to increase the 
transition temperature. There are many proteins that would be expected to 
do this, and the effect should be observable upon the introduction of 
simple peptides.

\section{Acknowledgments} 
We are grateful for many stimulating conversations with Caitlin Cornell 
and Sarah Keller. Useful exchanges with Sarah Veatch and Ilya Levental are 
also acknowledged with pleasure.

\end{document}